# Superposition of similar single-sided RTN processes and *1/f* noise

Giovanni Zanella

*Dipartimento di Fisica dell'Università di Padova and Istituto Nazionale di Fisica Nucleare, Sezione di Padova, via  Marzolo 8, 35131 Padova, Italy*

In this paper it is demonstrated that a *1/f* power spectrum appears in the process originated by the superposition of many similar single-sided RTN processes with the same relaxation time. The non-relaxed regime, the Gaussian nature  and the average periodicity of the resulting fluctuation, are responsible for the generation of  *1/f* noise, thanks to the coincidences of these RTN processes. The decomposition of the resulting fluctuation in a set of further single-sided RTN processes with a distribution of the relaxation times, permits us to demonstrate once more the generation of the *1/f* noise.

## 1.  Introduction

The main feature of  *1/f* noise is that its power spectrum increases with decreasing frequency *f* down to the lowest possible frequencies for conducting measurements. This noise is therefore spectrally *scaling*, that is, it is statistically identical to its transformation by contraction in time or another independent variable, followed by a corresponding change in intensity. This scaling property is typical of the *fractals* [1] and it reveals the presence of long-term correlations.

Despite great progress being made in *1/f* noise physics the source of these fluctuations remains unknown in the case of most systems and the problem remains largely unsolved in its generality.

 It would be too lengthy to report all references concerning *1/f* noise but an extensive bibliography has been drawn up in [2].

 The aim of this contribution is to furnish further focus and explanation, in respect to a previous work [3], to go back to the origin and the essence of the *1/f* noise.

  We shall see that a  *1/f* power spectrum appears in the resulting fluctuation from many, microscopic or macroscopic, similar single-sided RTN (random telegraph noise) processes with the same relaxation time, tanks to a surprising cooperative process.

In the following, we refer only to time phenomena.

## 2.  Role of the coincidences

Consider *N* similar two-state processes generated by the random presence, and by the absence, of a same physical  object (one electron, one photon, one car, etc). These processes can be represented by the two-level independent variables $u_1, u_2, u_3, ..., u_N$ having the same: amplitude *u*, probability *p* and relaxation time *t* (*the average time* *spent in a level before making a transition to the other or the inverse of the total rate of transitions back and forth in the process).*

The variable $u_i$ appears as a single-sided RTN process which ranges from the value zero to the value $u$.

Consider the number $c$ of *coincidences* which result in the superposition of the processes $u_i$. These coincidences correspond to the number of contemporary high levels during the sampling time $\Delta T$, while the entire measurement time is $T$.

Call $P(c)$ the probability of $c$ coincidences in the interval $\Delta T$.

If $U$ represents the summation of all the $u_i$ processes, we call $\Delta U$ the double-sided fluctuation of $U$ from the mean $\langle U \rangle$ and $\Delta c$ the corresponding double-sided fluctuation of $c$ from the mean $\langle c \rangle$.

The statistics which govern the probability $P(c)$ is binomial, where the processes $u_i$ generate $N$ independent trials. The memory of the processes $u_i$ does not compromise the final distribution of the coincidences, for it depends only from the probability $p$ and from $N$, although this memory is transmitted to the resulting process $\Delta U$.

## 2.1. *Biunique correspondence*

The particularity of the amplitudes $\Delta U$ is their biunique correspondence with the coincidences, thanks to the single-sided originating processes. In other words, the values of $U$ are discrete and they can be reached only through the eventuality of $c = U/u$ coincidences, while the fluctuation $\Delta U$ requires an increment $\Delta c = \Delta U/u$ from $\langle c \rangle$.

In this respect, we note that also in the simple case of RTN double-sided originating processes this biunique correspondence does not appear, due to the impossibility to go back to the number of coincidences causing the fluctuations.

## 2.2. *Rarity of the fluctuations*

In the case of a number sufficiently high of originating RTN processes, the biunique correspondence of $U$ with $c$ is responsible of the rarity of the fluctuations $\Delta U$ of great amplitude, in comparison to fluctuations of the same amplitude generated by random variables of continuous type. In fact, in this latter case, the same value of $\Delta U$ can be reached by various and different combinations of the generating processes and not by a unique possibility.

For example, supposing to have one hundred of RTN single-sided processes with $p = 0.5$ and $t = 1ms$, the probability of the coincidence of all hundred processes during $1ms$, corresponds, on average, to one possibility in about $4 \ 10^{16}$ years, when the retained age of the Universe is $2 \cdot 10^9$ years.

The rarity of the fluctuations explains also, as we shall see, the long-term memory.

## 2.3. Periodic structure of the average process

The crucial role of the coincidences is the locking of the originating process. This locking is responsible of the convergence of the resulting random process $\Delta U$ to an average periodic process. Indeed, the locking of $c$ processes requires the contemporary locking of the fluctuations related to $c$-1, $c$-2, … , $c$- ($c$-1) coincidences, permitting to the process $\Delta U$ to reach averagely a self-similar periodic structure.

The fundamental frequency of this average process is the inferior limit of the bandwidth of the process itself. On the contrary, the resulting fluctuations from originating processes different from single-sided RTN ones cannot converge to a periodic signal due to the lack of biunique correspondence between $c$ and $\Delta U$.

Fig.A and Fig.B display as the coincidences of five single-sided similar RTN processes work to create the resulting process $\Delta U$.

## 2.4. Waveform of the fluctuations

The fluctuation $\Delta U$, as we can see in Fig.A and Fig.B (page 12), show the following characteristics:
1. The average structure is self-similar "fractal" and reveals a self-organization process such as the generation of one-dimension sand-piles [4].
2. The amplitude of the fluctuations scales inversely with their frequency.
3. The duration (or the probability $p$) of the fluctuations at the same level of amplitude appears decreasing as their average frequency.
4. The waveform of $\Delta U$ is represented by symmetrical spikes whose sharpening naturally evolves with the number of the coincidences. This waveform is similar to the simulated signal in ref. [5] having a power spectrum proportional to $1/f$.

## 2.5. Gaussian distribution

When the number of the originating RTN processes is high, and the probability $p$ is not too small, the binomial distribution approaches the Gaussian distribution. In any case, the Gaussian distribution can be reached also thanks to the *central limit theorem*. Therefore, if $s_{\Delta c}^2$ is the variance of the distribution, it results

$$P(\Delta c) \propto e^{\frac{\Delta c^2}{2 s_{\Delta c}^2}} \, , \qquad (1)$$

where $P(\Delta c)$ is put in the place of $P(c)$ for the symmetrical distribution of $c$ in respect to $\langle c \rangle$. The same distribution occurs for the probability $P(\Delta U)$, being $U = cu$, so

$$P\left(\Delta U\right) \propto e^{-\frac{\Delta U^2}{2s_{\Delta U}^2}} \ .\qquad\qquad(2)$$

If we look to the fluctuation $\Delta U$ of discrete amplitude $\Delta U_1$, its power $\Delta U_1{}^2$ will be

$$\Delta U_1{}^2 = 2s_{\Delta U}^2\left[\ln\frac{1}{P(\Delta U_1)}+const\right].\qquad\qquad(3)$$

In particular, when $\Delta U_1{}^2 = 0$, $P(\Delta U_1) = P(\Delta U)_{max}$ , so

$$const = \ln P(\Delta U)_{max}\qquad,\qquad\qquad(4)$$

and

$$\Delta U_1{}^2 = 2s_{\Delta U}^2\ln\frac{P(\Delta U)_{max}}{P(\Delta U_1)}\qquad.\qquad\qquad(5)$$

Under the experimental point of view, we conduct measurements only during a finite time $T$. So, on $T/\Delta T$ alternatives there can be various values of power which have the same probability $P(\Delta U_i)$ (ratio of the number of the favourable trials $n_i$ with the number of the possible trials $n_{tot} = T/\Delta T$). Hence, $\Delta U_1{}^2$ represents *the average power of the events which have the same ratio $n_1 / n_{tot} = P(\Delta U_1)$ in the time $T$.*

## 3. Relaxed and non-relaxed regime

In general, the relaxation time of a process is the time $t$ necessary to its autocorrelation function to decrease of a factor $1/e$ .

Hence, a process $U$ can result relaxed or non-relaxed depending on the measurement time $T$. Indeed, the process appears relaxed if the measurement time exceeds its relaxation time ($T > t$), otherwise the process is non-relaxed ($T < t$).

We recall that, in the case of generating RTN processes, $t$ is also *the inverse of the total rate of transitions back and forth in the process* [2]. Therefore, if a fluctuation of discrete amplitude ?$U$ does not reappear within the measurement time $T$, the process is non-relaxed because $T < t$.

Considering the example of Sec. 2.2, and supposing ten generating RTN processes, the coincidence of all ten these processes will appear averagely every millisecond. So, with $T > 1ms$ the process will be relaxed, but with $T < 1ms$ the process will become non-relaxed, while with one hundred of processes the resulting process appears non-relaxed for any experimental $T$.

On the contrary, when a process results from the summation of many independent continuous random variables its regime is practically relaxed for any experimental $T$, due to the high statistics of the fluctuations and to the lack of memory of the process.

### 3.1. *Relaxation and memory*

The relaxation time of a process is accepted to represent its memory time. Therefore, if a fluctuation of discrete amplitude $?U$ does not reappear within the measurement time $T$, the memory of this fluctuation is not lost. Hence, the process $?U$ generated by many RTN processes, with the same relaxation time, shows a long-term memory, due to the presence of the rare fluctuations.

The long-term memory results also from the possibility to decompose the $?U$ process in a set of single-sided processes with a distribution of relaxation times (Sec.6).

Instead, a process without memory (white noise) will be always relaxed for any experimental $T$, while a relaxed process not necessarily implies a process without memory (Sec.4).

### 3.2. *Non-relaxation and bandwidth*

The sampling time $\Delta T$ limits the bandwidth of the process $\Delta U$ at $f_s = 1/(2?T)$ and the measurement time $T$ does not permit the detection of spectral components with a frequency lower than $f_T = 1/(2T)$.

Therefore, the measurement time $T$, and the sampling time $\Delta T$, filter the spectral components of a process $\Delta U$.

In addition to the limits introduced by $\Delta T$ and $T$, the bandwidth of the process is also determined by the bandwidth $f_2 < f < f_1$ of the measuring device, which interferes in various ways with the interval $f_T$ , $f_s$. For instance, if $f_2$ is the inferior limit of the bandwidth, it is necessary have $f_T \le f_2$, or $1/(2T) \le f_2$, to maintain this bandwidth.

In conclusion, to be sure to conduct measurements in the non-relaxed regime, it must be $T < t$, then $1/T > t^{-1}$, namely

$$t^{-1} < 2f_T = 1/T. \qquad (6)$$

## 4. The power spectrum of the fluctuations

We are interested to determine the power spectrum of the fluctuations $\Delta U$ of $U$. Therefore, it is useful to analyse these fluctuations in the relaxed regime and to see the difference with the non-relaxed regime.

### 4.1. *Relaxed regime*

In general, we suppose a measurement time $T \to \infty$ to guarantee the complete relaxation of the autocorrelation function $\psi_U(q)$ of $U$.

Being $U = \sum_{i=1}^{N} u_i$ its we have

$$\boldsymbol{y}_U(\boldsymbol{q}) = \lim_{T \to \infty} \frac{1}{T} \sum_{i=1}^{N} \sum_{j=1}^{N} \int_{-T/2}^{T/2} u_i(t)\, u_j(t-\boldsymbol{q})\, dt \quad . \tag{7}$$

In Eq.(7) the terms with $i=j$ correspond to the autocorrelation functions of the single processes $u_i$, while the other terms ($i \neq j$) are responsible for the cross-correlation contributions to $U$. In general we can write

$$\boldsymbol{y}_U(\boldsymbol{q}) = N\boldsymbol{y}_a(\boldsymbol{q}) + N(N-1)\boldsymbol{y}_c(\boldsymbol{q}) \quad , \tag{8}$$

where $\boldsymbol{y}_a(\boldsymbol{q})$ denotes the identical autocorrelation function for each involved $u_i$ process and $\boldsymbol{y}_c(\boldsymbol{q})$ the identical cross-correlation function between any two different $u_i$ processes.

As the processes $u_i$ are uncorrelated, but non-orthogonal

$$\boldsymbol{y}_c(\boldsymbol{q}) = \left\langle u_i u_j \right\rangle = \left\langle u_i \right\rangle^2 \neq 0 \qquad (\text{for } i \neq j) \,. \tag{9}$$

Therefore the average power of $U$ will be

$$\left\langle U^2 \right\rangle = \boldsymbol{y}_U(0) = N\boldsymbol{y}_a(0) + N(N-1)\boldsymbol{y}_c(0) \quad . \tag{10}$$

Now

$$\boldsymbol{y}_a(0) = \left\langle u_i^2 \right\rangle = \left\langle u_i \right\rangle^2 + \boldsymbol{s}_{u_i}^2 \quad \text{and} \quad \boldsymbol{y}_c(0) = \boldsymbol{y}_c(\boldsymbol{q}) = \left\langle u_i u_j \right\rangle = \left\langle u_i \right\rangle^2 \quad , \tag{11}$$

where $\boldsymbol{s}_{u_i}^2$ denotes the variance of the variable $u_i$.

Then

$$\left\langle U^2 \right\rangle = N^2 \left\langle u_i \right\rangle^2 + N\boldsymbol{s}_{u_i}^2 = \left\langle U \right\rangle^2 + \boldsymbol{s}_U^2 \quad , \tag{12}$$

being $\left\langle U \right\rangle^2 = N^2 \left\langle u_i \right\rangle^2$ , $\boldsymbol{s}_U^2 = N\boldsymbol{s}_{u_i}^2$ and $\boldsymbol{s}_U^2$ representing the average power of the fluctuations $\Delta U$ .

We can verify that the average power due to the dc level, namely the term $N^2 \left\langle u_i \right\rangle^2$ of Eq.(12), is the same of that due to the mean of the Gaussian distribution of the amplitudes of $U$. In fact, $\left\langle u_i \right\rangle = pu$ , so $N^2 \left\langle u_i \right\rangle^2 = (Npu)^2$ which is also the squared of the mean of the Gaussian distribution.

Thanks to the *Wiener-Khintchine theorem* the power spectrum $S(\boldsymbol{w})_U$ of $U$ will be the Fourier transform of $\boldsymbol{y}_U(\boldsymbol{q})$, that is

$$\begin{aligned} S(\boldsymbol{w})_U &= N\,\Psi_a(\boldsymbol{w}) + N(N-1)\,\Psi_c(\boldsymbol{w}) = \\ &N\,S(\boldsymbol{w})_u + N(N-1)\left\langle u_i \right\rangle^2 \boldsymbol{d}(\boldsymbol{w}) \end{aligned} \quad , \tag{13}$$

where $\Psi_a(\boldsymbol{w}) = S(\boldsymbol{w})_u$ and $\Psi_c(\boldsymbol{w}) = \langle u_i \rangle^2 \boldsymbol{d}(\boldsymbol{w})$ are the Fourier transform of $\boldsymbol{y}_a(\boldsymbol{q})$ and $\boldsymbol{y}_c(\boldsymbol{q})$.

Eq.(13) confirms that, in the frequency domain, the dc components of the originating processes introduce still a dc component in the resulting process, while for any other frequency the power spectrum results from the summation of the single power spectra of the processes $u_i$.

Hence, being Lorentzian the power spectrum of the originating RTN processes in a relaxed regime, the power spectrum of the resulting process $?U$, in the same regime, will be still Lorentzian.

As a consequence, the power spectrum of $\Delta U$ will can be of *1/f* type only during a non-relaxed regime.

### 4.2. *Non-relaxed regime*

This regime appears when the measurement time $T$ does not permit to the fluctuations $\Delta U$, with a average frequency lower than $f_T = 1/(2T)$, to appear with repetition. Hence, the autocorrelation function does not relax and the process does not loss its memory within $T$. Therefore, the resulting process of many similar originating RTN processes will be always non-relaxed within experimental measurement times and its power spectrum will be different from a Lorentzian.

## 5. *1/f* power spectrum and Gaussian distribution

We have seen that the process $\Delta U$, derived from the summation of many independent RTN processes, follows a Gaussian distribution and it is almost always non-relaxed, within experimental measurement times (or equivalent bandwidth) due to the rare coincidences. On the other hand, the *law of large numbers* allows us to introduce an average frequency of the coincidences proportional to their probabilities, thereby enabling us to substitute, in the Eq.(5), the discrete probability $P(\Delta U)$ with the discrete frequency $f = P(\Delta U)/\Delta T$.

In the frequency domain Eq. (5) becomes

$$\Delta U_1^{\ 2} = 2\boldsymbol{s}_{\Delta U}^{\ 2} \ln \frac{f_{max}}{f_1} \quad , \qquad (14)$$

where $\Delta U_1^{\ 2}$ denotes *the average power of all the fluctuations of frequency $f_1$* . Now, if the average structure of $\Delta U$ is periodic (Sec. 2.3), $\Delta U_1^{\ 2}$ represents the average power of the process $\Delta U$ in the period $1/f_1$. This periodic process $\Delta U_1$ has fundamental frequency $f_1$ with spectral components in the frequency range $f_1, f_{max}$ and its average power is just the summation of the average power of its spectral components [8].

Introducing the power spectrum $S_{\Delta U}(f)$, we have

$$\int_{f_1}^{f_{max}} S_{\Delta U}(f)\, df = 2 s_{\Delta U}^2 \ln \frac{f_{max}}{f_1} \quad , \qquad (15)$$

that is

$$S_{\Delta U}(f) = \frac{2 s_{\Delta U}^2}{f} \quad , \qquad (16)$$

where $S_{\Delta U}(f)$ represents the power spectrum of the spectral components necessary to produce the power $\Delta U_1^2$.

Thanks to the dependence of $S_{\Delta U}(f)$ on $s_{?U}^2$ it is possible the demonstration of the Hooge formula [3].

## 6. *1/f* power spectrum by summation of many similar RTN processes

Fig. 1A shows a simple example of fluctuation $\Delta U$, due a summation of similar RTN processes of amplitude $u$, where the maximum amplitude of the fluctuations is limited to $\Delta U = 5u$ and the process is viewed only on side of the fluctuations $\Delta U$.

We see in Fig.C that $\Delta U$ can also be generated by an equivalent set of similar RTN processes of amplitude $u$ and with a relaxation time which scales as $\Delta U$.

Each of these processes has its own relaxation time $t_{?c}$, probability $p_{?c}$ (in the high level) and it is in biunique correspondence with $\Delta c$ of average frequency $f_{?c}$.

The relaxation time $t_{?c}$ obeys the relationship

$$\frac{1}{2 t_{\Delta c}} = \frac{p_{\Delta c}}{\Delta T} = f_{\Delta c} \qquad , \qquad (17)$$

indeed, $1/\Delta T$ represents the number of samples in a time-unit and $p_{?c}/\Delta T$ the frequency $f_{?c}$ of the events $\Delta c$, where $f_{?c} = 1/(2 t_{?c})$.

Therefore, the probability $p_{?c}$ will be

$$p_{\Delta c} = \frac{1}{2 t_{\Delta c}} \Delta T \propto \frac{1}{t_{\Delta c}} \qquad . \qquad (18)$$

Hence the power spectrum $S(?)$ of the resulting process $\Delta U$ will be

$$S(w) = \sum_{\Delta c} p_{\Delta c} u^2 \frac{4 t_{\Delta c}}{1 + w^2 t_{\Delta c}^2} \propto \sum_{\Delta c} \frac{1}{t_{\Delta c}} \frac{4 t_{\Delta c}}{1 + w^2 t_{\Delta c}^2} \qquad , \qquad (19)$$

Indeed, $p_{\Delta u}u^2$ represents the average power, or the variance, of a RTN process as those of Fig.C, since they are single-sided.

In the case of a superposition of many RTN processes with a continuous distribution of the relaxation times, $S(w)$ has to be averaged over this distribution with a function weight $p(t)\, dt$ which includes the contribution to the variance of processes whose relaxation time lies in the interval from $t$ to $t + dt$ . Supposing we have $p(t) \propto 1/t$ in some interval $t_2^{-1} << t^{-1} << t_1^{-1}$ and zero outside this interval, one has $S(w) \propto 1/w$ in the frequency range $t_2^{-1} << w << t_1^{-1}$ [1] [5]. Therefore

$$S(w) \propto \int_{t_2}^{t_1} \left(\frac{1}{t}\right)\frac{t}{1+w^2 t^2}\, dt \propto \frac{1}{w} \ . \qquad (20)$$

This result has been possible thanks to the following conditions:

1. The fluctuations $\Delta U$ are originated by RTN processes of the same amplitude, so they can be decomposed by one set of RTN processes of the same amplitude with a distribution of relaxation times.
2. The non-relaxed regime is guaranteed by the condition $t_2^{-1} << 2\pi f$, namely $t_2^{-1} < 2f$ [see Eq.(6)] and this requires a measurement time $T=1/(2f)$.

Now, these conditions can be respected if the fluctuations $\Delta U$ are just generated by the coincidences of many originating RTN processes, with the same amplitude and relaxation time.

It is evident in Fig.C that the long-term memory results also by the presence, in the equivalent set of RTN processes, of a relaxation time increasing with the decrease of the average frequency of $\Delta U$.

At least, the *1/f* power spectrum conducts to a Gaussian distribution of the fluctuations $\Delta U$ through the reverse of the path described in Sec.5. Indeed, the average power of the fluctuation $\Delta U$ in the frequency range $f_{max}, f_1$ will be

$$\Delta U_1^2 = \int_{f_1}^{f_{max}} S_{\Delta U}(f)\, df \propto \ln\frac{f_{max}}{f_1} \qquad (21)$$

that is

$$f_1 = f_{max}\ e^{-\frac{\Delta U_1^2}{const}} \ . \qquad (22)$$

Eq. (22) describes a periodic Gaussian process where $\Delta U_1^2$ represents the average power in the period $1/f_1$ .

The Gaussian distribution of the *1/f* noise was found in measurements performed by R.F. Voss on different solid state devices [8]. In particular R.F. Voss tested the correlation between Gaussian behaviour and *1/f* noise observed in a sufficiently pure form.

## 7. Serial 1/f noise

A serial single-sided RTN process is very common in nature, for example the sequential flow, along the same path, of identical objects such as electrons, photons, cars, etc.

If $\Delta t$ is the minimum time necessary to detect one object which flows through a cross-section of the path, $N = T/\Delta T$ represents the number of attempts in the time $T$.

If $p$ is the probability of detecting one of these objects in time $\Delta T$ then the probability $P(n)$ to detect $n$ objects in the time $T$ follows the binomial distribution.

In particular, if the objects have the same velocity, the time $T$ can be viewed as space $S$.

When $N$ is large and the probability $p$ is not very small, $P(n)$ will follow the Gaussian distribution, so

$$P(n) \propto \exp\left[-\frac{(n-\langle n \rangle)^2}{2s_n^2}\right] \quad , \qquad (23)$$

where $\langle n \rangle$ denotes the mean value of $n$ and $s_n$ the standard deviation of the distribution.

Solving Eq. (23) in the unknown $\Delta n^2 = (n-<n>)^2$, we obtain

$$\Delta n^2 = 2s_{\Delta n}^2 \ln \frac{P(\Delta n)_{\max}}{P(\Delta n)} \quad , \qquad (24)$$

where $P(\Delta n) = P(n)$, $s_{\Delta n} = s_n$ and $P(\Delta n)_{\max} = \dfrac{1}{s_{\Delta n}\sqrt{2p}}$.

Thanks to the *law of large numbers*, the mean frequency $f$ of the fluctuations $\Delta n$ is proportional to their probabilities and therefore we can substitute $P(n)$ with $f$. In fact $f = P(\Delta n) / \Delta T$ and $f_{max} = P(\Delta n)_{\max}/\Delta T$.

$\Delta n^2$ represents the average power of the fluctuations $\Delta n$ of average frequency $f$ which spectral components lie in the range $f, f_{\max}$.

Thus, as in Sec.6, we introduce the power spectrum $S_{\Delta n}(f)$ such that

$$\int_{f}^{f_{\max}} S_{\Delta n}(f)\,df = \Delta n^2 = 2s_{\Delta n}^2 \ln \frac{f_{\max}}{f} \quad . \qquad (25)$$

so

$$S_{\Delta n}(f) = 2s_{\Delta n}^2/f \quad . \qquad (26)$$

## 8. Conclusions

In this paper it has been demonstrated that *1/f* noise is produced when a superposition exists of many similar single-sided RTN processes. This result is possible thanks:

- the non-relaxed regime of the resulting process $\Delta U$, guaranteed by an measurement time shorter than the relaxation time, or by the equivalent bandwidth;
- the Gaussian distribution of the fluctuations $\Delta U$;
- the periodic average structure of $\Delta U$.

The resulting process $\Delta U$ can be also decomposed by an equivalent set of RTN processes with relaxation time scaling as the amplitude of the fluctuations $\Delta U$. This set of RTN processes, with a distribution of relaxation times, permits to operate on the frequency range of a the non-relaxed regime $\Delta U$ and to find an $1/f$ power spectrum. In this case the Gaussian distribution derives as a consequence of an exact $1/f$ noise.

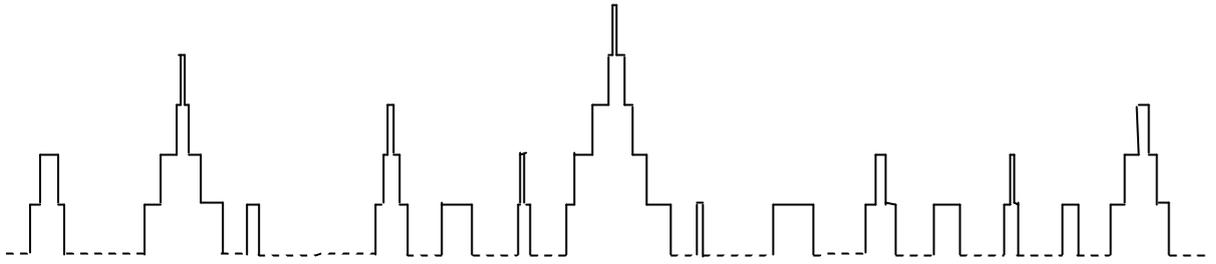

A

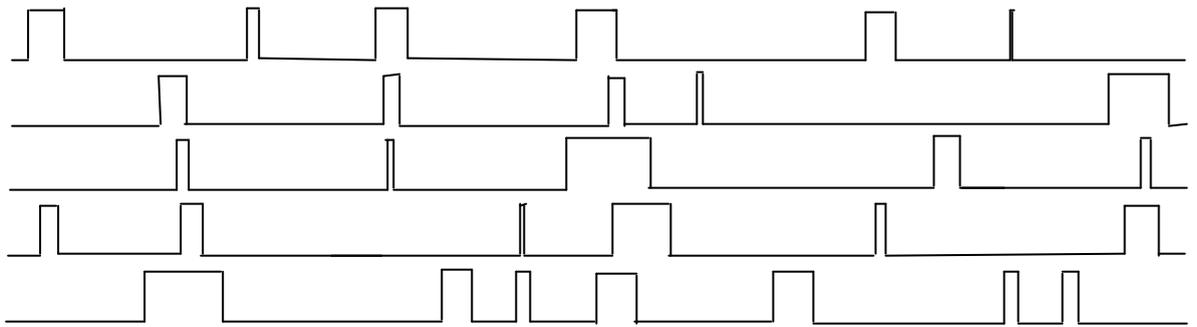

B

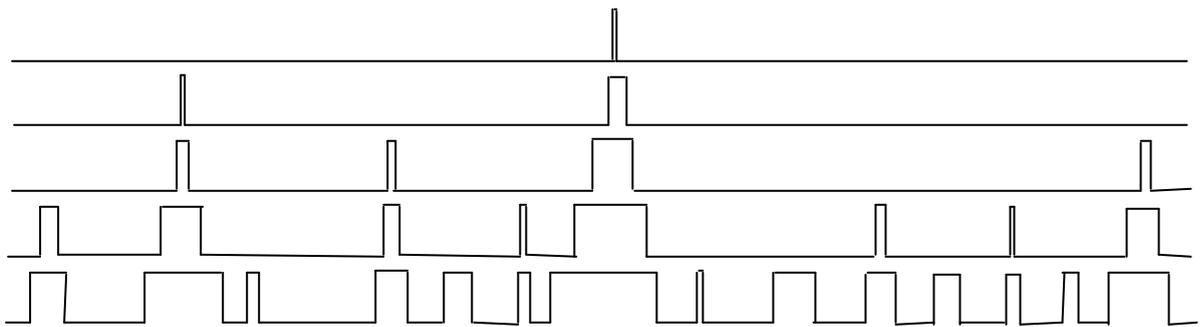

C